\documentclass[english,preprint,onecolumn,showpacs,preprintnumbers,amsmath,amssymb,showkeys]{revtex4-2}

\usepackage{chemformula} % Formula subscripts using \ch{}
\usepackage[T1]{fontenc} % Use modern font encodings
\usepackage{bm}
\usepackage[normalem]{ulem}
\usepackage[T1]{fontenc}
\usepackage{color}
\usepackage{units}
\usepackage{amssymb}
\usepackage{graphicx}
\usepackage{esint}
\usepackage{bm}
\usepackage{natbib}
\usepackage{ulem}
\usepackage{babel}
\usepackage[colorlinks=true, citecolor=red]{hyperref}

\begin{document}

\title{Coexisting Massive and Massless Dirac Fermions in Moir\'{e}-Reconstructed Bilayer Graphene}
\author{Mohit Kumar Jat$^1$, Kenji Watanabe$^2$, Takashi Taniguchi$^3$}
\author{Aveek Bid$^1$}
\email{aveek@iisc.ac.in}
\affiliation{$^1$Department of Physics, Indian Institute of Science, Bangalore 560012, India \\
	$^2$ Research Center for Electronic and Optical Materials, National Institute for Materials Science, 1-1 Namiki, Tsukuba 305-0044, Japan\\
	$^3$ Research Center for Materials Nanoarchitectonics, National Institute for Materials Science,  1-1 Namiki, Tsukuba 305-0044, Japan\\}

\setlength {\marginparwidth }{2cm}

\begin{abstract}
We report the emergence of massless Dirac fermions in moir\'{e}-reconstructed bands of bilayer graphene (BLG) aligned with hexagonal boron nitride (hBN). Magnetotransport measurements reveal that while the primary BLG band retains a parabolic dispersion with a Berry phase of $2\pi$, the moir\'{e}-induced secondary bands at $n/n_0 = \pm 4$ host chiral massless quasiparticles with a Berry phase $\pi$ and a Fermi velocity $v_m \approx 3.6 \times 10^5 \mathrm{m s^{-1}}$. This transition from massive to massless carriers arises from topological band reconstruction driven by the hBN moir\'{e} potential. Our results demonstrate that moir\'{e} engineering in BLG/hBN offers a powerful route to tune band topology and realize coexisting Dirac and massive fermions within a single crystalline platform.
\end{abstract}

\maketitle

\section{Introduction}
Realizing massless Dirac fermions in a system that is nominally parabolic, such as bilayer graphene (BLG), offers an exciting opportunity to control topological band structure and quantum transport at will. In pristine BLG, low-energy carriers possess a finite mass and exhibit a Berry phase of $2\pi$~\cite{Novoselov2006, PhysRevLett.96.086805}, in contrast to the linear, massless dispersion of monolayer graphene~\cite{Novoselov2005, Zhang2005}. Engineering a transition between these two regimes-massive and massless-within a single material system provides a platform for exploring novel transport phenomena, tunable topological states, and emergent correlated phases.

When BLG is placed on hexagonal boron nitride (hBN) with a small twist angle, the resulting moir\'{e} potential introduces a long-wavelength periodic modulation that profoundly alters the electronic structure~\cite{PhysRevB.90.155406, PhysRevB.88.205418, Jat2024, doi:10.1126/science.adh3499,Pantaleon2021}. This superlattice potential can generate secondary minibands, and Hofstadter-like fractal spectra~\cite{PhysRevB.94.045442, Wang2015,Jeong2024}. Previous transport and thermodynamic~\cite{ doi:10.1126/science.adh3499} studies of BLG/hBN moir\'{e} have reported superlattice features, carrier scattering~\cite{Jat2024}, gapless minibands~\cite{Kim2018}, and interaction-driven phases~\cite{Jeong2024, doi:10.1126/science.aan8458}. In addition, secondary Dirac points (SDP) have also been observed in single-layer graphene/hBN moir\'{e} heterostructures~\cite{Yankowitz2012, Ponomarenko2013,doi:10.1126/science.1237240,Kim2018}. However, in the BLG, the topological nature of these minibands and their corresponding mass renormalization have not been quantitatively established. In particular, a clear experimental determination of whether these moir\'{e}-induced minibands host massive or massless quasiparticles, and how their Berry phases evolve, has been lacking.

The physics of these moir\'{e} minibands can be qualitatively understood within the framework of the Diophantine equation~\cite{PhysRevLett.49.405},
\begin{equation}
    \frac{n}{n_0} = t \left( \frac{\phi}{\phi_0} \right) + s,
\end{equation}
where $n_0$ is the carrier density corresponding to one electron per moir\'{e} unit cell, $\phi/\phi_0$ is the normalized magnetic flux, and the integers $(t, s)$ encode the Chern number and miniband index, respectively~\cite{PhysRevLett.45.494,TillSchlosser_1996}. In the presence of a strong moir\'{e} potential, states with non-zero $s$ emerge, giving rise to distinct topological minibands whose properties can be probed via magnetotransport~\cite{doi:10.1126/science.aan8458, Dean2013, PhysRevLett.86.147, doi:10.1126/science.1237240}.

Here we demonstrate experimentally that the moir\'{e} potential in aligned BLG/hBN transforms the parent massive dispersion into a Dirac-like band with a $\pi$ Berry phase-establishing a tunable platform for hybrid massive–massless fermion physics. The moir\'{e}-induced bands shows a reduced Fermi velocity compared to the pristine graphene, highlighting that the hBN moir\'{e} leads to subtle band flattening effects. Recent observations of emergent phases like Chern insulating states and charge density wave in rhombohedral graphene/hBN moir\'{e} heterostructure highlight the need for a detailed study on the role of hBN moir\'{e} in modifying the band topology of 2D materials~\cite{Lu2024, Chen2020, Wang2024,chsq-ndzs}. Our present study further emphasizes the influence of the hBN moir\'{e} potential in BLG/hBN, which drives a transition from a parabolic to a Dirac-like band topology.

\section{Results and Discussion}

\begin{figure}[t]
	\includegraphics[width=\columnwidth]{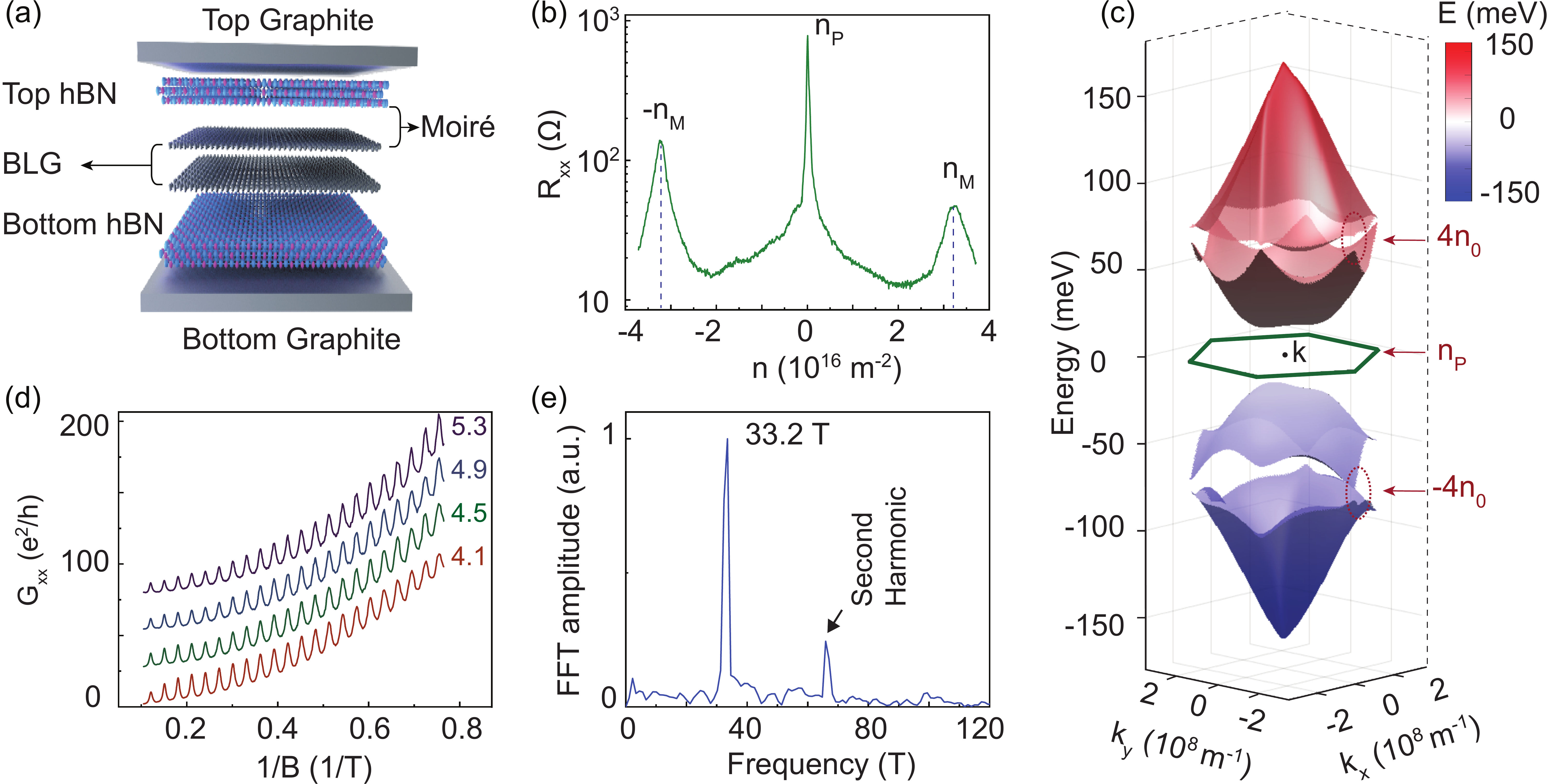}
	\small
	\caption{\textbf{Device structure and basic characterization of the moiré BLG/hBN heterostructure.}
		(a) Schematic cross-section of the hBN/BLG/hBN stack showing the moiré pattern formed between BLG and the top hBN layer.
		(b) Longitudinal resistance $R_{xx}$ versus carrier density $n$ measured at $T = 2~\mathrm{K}$, showing satellite peaks at $n_M = \pm3.2\times10^{16}~\mathrm{m^{-2}} = 4n_0$ (dashed blue lines) arising from moiré miniband formation.
		(c) Three-dimensional schematic of the first and second electron and hole bands near the $K$-valley, illustrating the moiré-induced reconstruction of the BLG dispersion. (d) Brown–Zak magnetoconductance oscillations as a function of inverse magnetic field $1/B$ for various carrier densities (labels in units of $10^{16}~\mathrm{m^{-2}}$) measured at $T = 100~\mathrm{K}$.
		(e) Fast-Fourier transform of the oscillations at $n = 4.1\times10^{16}~\mathrm{m^{-2}}$, revealing a single prominent peak at $f_{\mathrm{BZ}} = 33.2~\mathrm{T}$, {and its weak second harmonic peak at $66.4$~T} corresponding to the moiré periodicity {of $12$~nm}.
	}
	\label{fig:fig1}
\end{figure}

High-mobility dual graphite-gated heterostructures of hBN/BLG/hBN are fabricated using the dry transfer technique~\cite{Pizzocchero2016,wang2013one} (Section~S1 {of Supplemental Material (SM)~
\footnote{The Supplemental Material contains the details on device characterization, data on two additional devices, and details of data analysis.}) }with the layer schematic shown in Fig.~\ref{fig:fig1}(a). The top hBN was aligned with the BLG to within $<1^\circ$ to create a moir\'{e} superlattice, while the bottom hBN was intentionally misaligned by $\sim 15^\circ$, ensuring that only the top moir\'{e} superlattice dominates the electronic reconstruction. The device is dry-etched into a Hall bar geometry to avoid mixing of longitudinal and Hall voltages~\cite{DivyaOMR}. Dual graphite gates allow independent control over the carrier density $n$ and the effective electric field $D$.

In the main text, we present the data for device \textbf{D1}, while additional results from devices ({\textbf{D2 and D3}}) are provided in Section S4 of SM~\cite{Note1}. The devices typically exhibit mobilities in the range of $(1-2) \times 10^5~\mathrm{cm^2 V^{-1} s^{-1}}$ at 2~K, confirming exceptional crystalline quality and low disorder. Fig.~\ref{fig:fig1}(b) shows the zero-field longitudinal resistance ${R_{xx}(B=0)}$ as a function of carrier density ($n$) at $T=2$~$\mathrm{K}$. In addition to the primary peak at charge neutrality  $n_P$ ($n=0$), distinct satellite peaks appear at the carrier densities $n_M=\pm 3.2\times10^{16}~\mathrm{m}^{-2}  = 4n_0$ as a consequence of moir\'{e}-induced band splitting in the electron and hole bands~\cite{Ponomarenko2013, Jat2024}. $n_0=2/(\sqrt{3}\lambda^2)$ is the filling of each moir\'{e} unit cell of superlattice wavelength $\lambda$, and the factor $4$ arises from the two-fold valley and spin degeneracies of BLG. From $n_M$, the moir\'{e} wavelength is extracted as $\lambda = 12$~nm, corresponding to a twist angle $\theta = 0.60^\circ$  between the BLG and the top hBN layer (Section~S2 of SM~\cite{Note1}). {Devices \textbf{D2} and \textbf{D3} have twist angles of $0.55^\circ$ and $0^\circ$, corresponding to superlattice wavelengths of $12.4$~nm and $14$~nm, respectively (Section S2 of SM~\cite{Note1})}.

Fig~\ref{fig:fig1}(c) shows the electronic dispersion relation of the first and second electron and hole bands at the $K$ valley. Along with the primary bands, one obtains moir\'{e}-reconstructed secondary bands appearing at carrier densities $\pm 4n_0$. The  Brillouin zone corresponding to moir\'{e} superlattice is highlighted by the green hexagon.

Strong Brown-Zak (BZ) oscillations~\cite{Jat202Natcomm, doi:10.1126/science.aal3357,doi:10.1073/pnas.1804572115,doi:10.1126/sciadv.aay8897} appear in the magnetoconductivity $G_{xx}=R_{xx}/(R_{xx}^2+R_{xy}^2)$ measured at $T = 100$~K (Fig.~\ref{fig:fig1}(d)). The oscillations exhibit a single Fourier peak at $f_{BZ} = 33.2$~T  {and its weak second harmonic peak at $66.4$~T} (Fig.~\ref{fig:fig1}(e)), consistent with the same moir\'{e} periodicity (Section~S3 of SM~\cite{Note1}).  The presence of these oscillations confirms high structural uniformity and a well-defined moir\'{e} potential landscape.

In the magnetotransport data (Fig.~\ref{fig:fig2}(a)), two distinct Landau fan patterns emerge-one from the primary BL band ($n/n_0=0$) and another from the moir\'{e}-induced secondary bands ($n/n_0=\pm4$). {The horizontal low-resistance features correspond to Brown-Zak oscillations at integer values of $\phi_0/\phi = q$ ($q \in \mathbb{Z}$)~\cite{doi:10.1126/science.aal3357,doi:10.1073/pnas.1804572115,Jat2024}.}
Fig.~\ref{fig:fig2}(b) shows the magnetoresistance oscillations of the primary band as a function of the Landau level (LL) filling factor $\nu_{P} = nh/eB$. At low magnetic fields, the resistance minima occur at $\nu_{P} = 4m$ ($m \in \mathbb{Z}$), reflecting the four-fold spin--valley degeneracy characteristic of massive quasiparticles in bilayer graphene~\cite{Novoselov2006, PhysRevLett.96.086805, McCann_2013}. A line cut of the data at $B=1$~T (Fig.~\ref{fig:fig2}(c))  reveals this periodicity.

The Landau fans emanating from the secondary miniband correspond to $s=4$ in the Diophantine framework, with $t=\nu_M = {(n-4n_0)h}/{eB}$. The resistance minima now occur at $\nu_M = 4(m+1/2)$ (Fig.~\ref{fig:fig2}(d -- e)). A half-integer shift in the Landau-level index sequence at $s = \pm4$ directly signals the emergence of massless Dirac quasiparticles.~\cite{Novoselov2005, Zhang2005}. Measurements on {additional devices, \textbf{D2} and \textbf{D3}}, reproduce these observations (Section~S4 of SM~\cite{Note1}). These results establish that the primary band ($t = 4m$, $s=0$) and the secondary band ($t = 4m+2$, $s=4$) have very different band topologies, ~{robust across BLG–hBN systems with near-zero alignment ($<1^\circ$).}

To quantify the carrier dynamics in the moir\'{e} minibands, we extracted the effective masses from the temperature dependence of the Shubnikov de Haas (SdH) oscillation amplitudes. In two-dimensional systems, $m^*$ is commonly modeled as an empirical function of carrier density,~\cite{Novoselov2005, Tiwari2022}
\begin{equation}
    \frac{m^*}{m_e} = A\!\left|\frac{n}{n_0}\right|^{\!\alpha},
\end{equation}
where $A$ and $\alpha$ are fitting parameters and $m_e$ is the free-electron mass. This dependence yields a corresponding dispersion relation (Section~S5 of SM~\cite{Note1}):
\begin{equation}
E(k) = E_0 + \frac{\hbar^2}{2A m_e (1 - \alpha)}
\left(\frac{4\pi n_0}{g_s g_v}\right)^{\!\alpha} k^{2(1-\alpha)}.
\label{eq:dispersion}
\end{equation}
Thus, the power-law exponent $\alpha$ directly reveals the curvature of the underlying energy–momentum relation.

Fig.~\ref{fig:fig3}(a) shows the temperature dependence of $R_{xx}$ versus $1/B$ at
$n = 4.05 \times 10^{16}\,\mathrm{m^{-2}}$ ($n/n_0 = 5.06$). Two distinct oscillation frequencies appear in the FFT (Fig.~\ref{fig:fig3}(b)), originating from the primary and secondary bands. Their density dependence (Fig.~\ref{fig:fig3}(c)) follows
$B_F^P = |n|h/g_s g_v e$ and $B_F^M = |n - 4n_0|h/g_s g_v e$, consistent with $g_s g_v = 4$ ($g_s$: spin degeneracy and $g_v$: valley degeneracy).  The separated frequencies enable band-resolved analysis using FFT filtering (Section~S6 of SM~\cite{Note1}). The observation of $g_s g_v = 4$ further indicates that alignment-induced valley splitting does not play a significant role in the analyzed minibands.

The thermal damping of the SdH oscillations was fitted to the Lifshitz–Kosevich (LK) formalism:~\cite {lifshitz1956theory,shoenberg1984magnetic}
\begin{equation}
\Delta R_{xx} \propto \frac{2\pi^2 k_B T m^*/\hbar e B}{\sinh(2\pi^2 k_B T m^*/\hbar e B)}.
\label{eq:LK}
\end{equation}
Fig.~\ref{fig:fig3}(d) displays the temperature dependence of the oscillation amplitude, and LK fits yield $m^*_P \approx 0.035\,m_e$ for the primary band-nearly constant across densities (Fig.~\ref{fig:fig3}(e)), confirming its parabolic dispersion.

In contrast, the secondary moir\'{e} band exhibits a strong density dependence (Fig.~\ref{fig:fig3}(f)), well described by
\[
\frac{m^*_M}{m_e} = A_M \left|\frac{n - 4n_0}{n_0}\right|^{\alpha_M},
\]
with $A_M = 0.051$ and $\alpha_M = 0.48 \pm 0.02$. Substituting these into Eq.~\ref{eq:dispersion} gives $
E(k) = \hbar v_M k_M$,
$v_M = \hbar\sqrt{\pi n_0}/(A_M m_e) = 3.6 \times 10^5\,\mathrm{m\,s^{-1}}$,
where $k_M = \sqrt{|(n - 4n_0)|\pi}$. The resulting linear dispersion $E \propto k^{1.0 \pm 0.1}$ signifies the emergence of massless quasiparticles in the moir\'{e} minibands.
The one-third reduction in $v_M$ relative to pristine graphene ($v_F = 1\times10^6\,\mathrm{m\,s^{-1}}$)~\cite{RevModPhys.81.109} reflects moir\'{e}-induced band flattening and enhanced correlation susceptibility -- key signatures of topological reconstruction.

The Berry phase $\Phi_B = 2\pi\beta$ serves as a direct probe of band topology. {We extract $\Phi_B$ from an analysis of Shubnikov-de Haas oscillations, where the oscillating part of the longitudinal resistance follows $\Delta R_{xx} \propto \cos [2\pi ({B_F}/{B} + 1/2 + \beta)]$~\cite{Novoselov2005, Zhang2005}, with $B_F$ being the oscillation frequency. Assigning integer Landau indices $N$ to the minima of $R_{xx}$ yields the relation $N = {B_F}/{B} + \beta$. Linear fit of $N$ versus $1/B$ directly yields the intercept $\beta$ (Section S10 of SM~\cite{Note1}).} Fig.~\ref{fig:fig4} presents the Landau-level (LL) index $N$ plotted against $1/B$ for both the primary and secondary bands. Linear fits yield intercepts that unambiguously distinguish the two regimes: $\beta_P = 1 \pm 0.05$ ($\Phi_B = 2\pi$) for the primary band (Fig.~\ref{fig:fig4}(a)), consistent with massive bilayer carriers~\cite{Novoselov2006}, and $\beta_M = 0.5 \pm 0.05$ ($\Phi_B = \pi$) for the secondary band (Fig.~\ref{fig:fig4}(b)), characteristic of chiral Dirac fermions~\cite{Novoselov2005}. {Additionally, the temperature dependence of the longitudinal resistance at $n/n_0 =4$ (Section~S9 of SM~\cite{Note1}) shows the system to be gapless, establishing that the charge carriers at the electron side SDP are massless Dirac fermions.}

The direct correspondence between the Berry phase and the scaling of $m^*_M$ with carrier density consolidates our conclusion: the moir\'{e} potential in aligned BLG/hBN transforms the parent parabolic BLG bands into emergent Dirac-like minibands. These results establish a robust experimental link between topological reconstruction, Berry-phase evolution, and effective-mass renormalization in moir\'{e}-engineered bilayer graphene.

\section{Conclusion}

In summary, alignment-controlled BLG/hBN superlattices host coexisting massive and massless quasiparticles arising from moiré-driven topological band reconstruction. The emergent Dirac bands display linear dispersion, while the parent BLG bands remain parabolic. Our findings demonstrate that moiré superlattices introduce periodicity, which selectively hybridizes and linearizes the electronic bands, allowing for the coexistence of Dirac and massive quasiparticles within the same bilayer graphene system. This coexistence provides a unique platform to study the interplay between massless and massive fermions, leading to rich quantum transport behavior and tunable correlated phases.

In particular, the band crossings near the secondary Dirac points exhibit reduced Fermi velocities ($v_{\mathrm{M}} = 3.6\times 10^5~\mathrm{m\,s^{-1}}$), indicating moiré-induced band flattening. Recently, anomalous Chern insulating states have been reported in both rhombohedral pentalayer graphene/hBN~\cite{Lu2024} and rhombohedral trilayer graphene/hBN moiré~\cite{Chen2020, Wang2024} heterostructures, suggesting that these effects may be linked to hBN-moiré–induced modifications of the parent graphene’s electronic topology. This intriguing connection highlights a broader design principle: by engineering moiré alignment, one can realize tunable Dirac fermions and correlated topological phases within a simple bilayer geometry.
\\

\textbf{Acknowledgment}
	A.B. acknowledges funding from the U.S. Army DEVCOM Indo-Pacific (Project number: FA5209   22P0166) and ANRF, DST (Project number: SPR/2023/000185). M.K.J. acknowledges funding from the Prime Minister's Research Fellowship (PMRF), MHRD. K.W. and T.T. acknowledge support from the JSPS KAKENHI (Grant Numbers 21H05233 and 23H02052) and World Premier International Research Centre Initiative (WPI), MEXT, Japan.
\\

	\clearpage

	%% Figure 1
\clearpage

	%%% Figure 2
\begin{figure}[t]
	\includegraphics[width=\columnwidth]{Fig2.pdf}
	\caption{\small
		\textbf{Landau-fan diagram and quantum oscillations revealing moiré-induced Dirac fermions.}
		(a) 2-D map of $R_{xx}$ as a function of moir\'{e} band filling $n/n_{0}$ and magnetic field $B$, showing Landau fan patterns emerging from the primary band ($n/n_{0} = 0$) and secondary bands ($n/n_{0} = \pm 4$). The horizontal low-resistance features marked with white arrows are Brown-Zak oscillations ($\phi_0/\phi = q$, with $q \in \mathbb{Z}$). (b) 2-D map of $R_{xx}$ as a function of $B$ and filling fraction $\nu_\mathrm{P} = nh/eB$ of the primary band. White dashed lines mark prominent minima at $\nu_\mathrm{P} = 4m$ ($m \in \mathbb{Z}$), characteristic of bilayer graphene. (c) Line cut of panel (b) at $B = 1$~T, showing clear $R_{xx}$ minima at $\nu_\mathrm{P} = 4m$. (d) 2-D map of $R_{xx}$ as a function of $B$ and the effective filling fraction of secondary band $\nu_\mathrm{M} = (n - 4n_{0})h/eB$. White dashed lines highlight minima at $\nu_\mathrm{M} = 4m + 2$, indicative of Dirac-like Landau level structure. (e) Line cut of panel (d) at $B = 4.5$~T, showing clear $R_{xx}$ minima at $\nu_\mathrm{M} = 4m + 2$. The color scales in panels (a), (b), and (d) are identical.}
	\label{fig:fig2}
\end{figure}

\clearpage

	% figure 3
\begin{figure}[t]
	\includegraphics[width=\columnwidth]{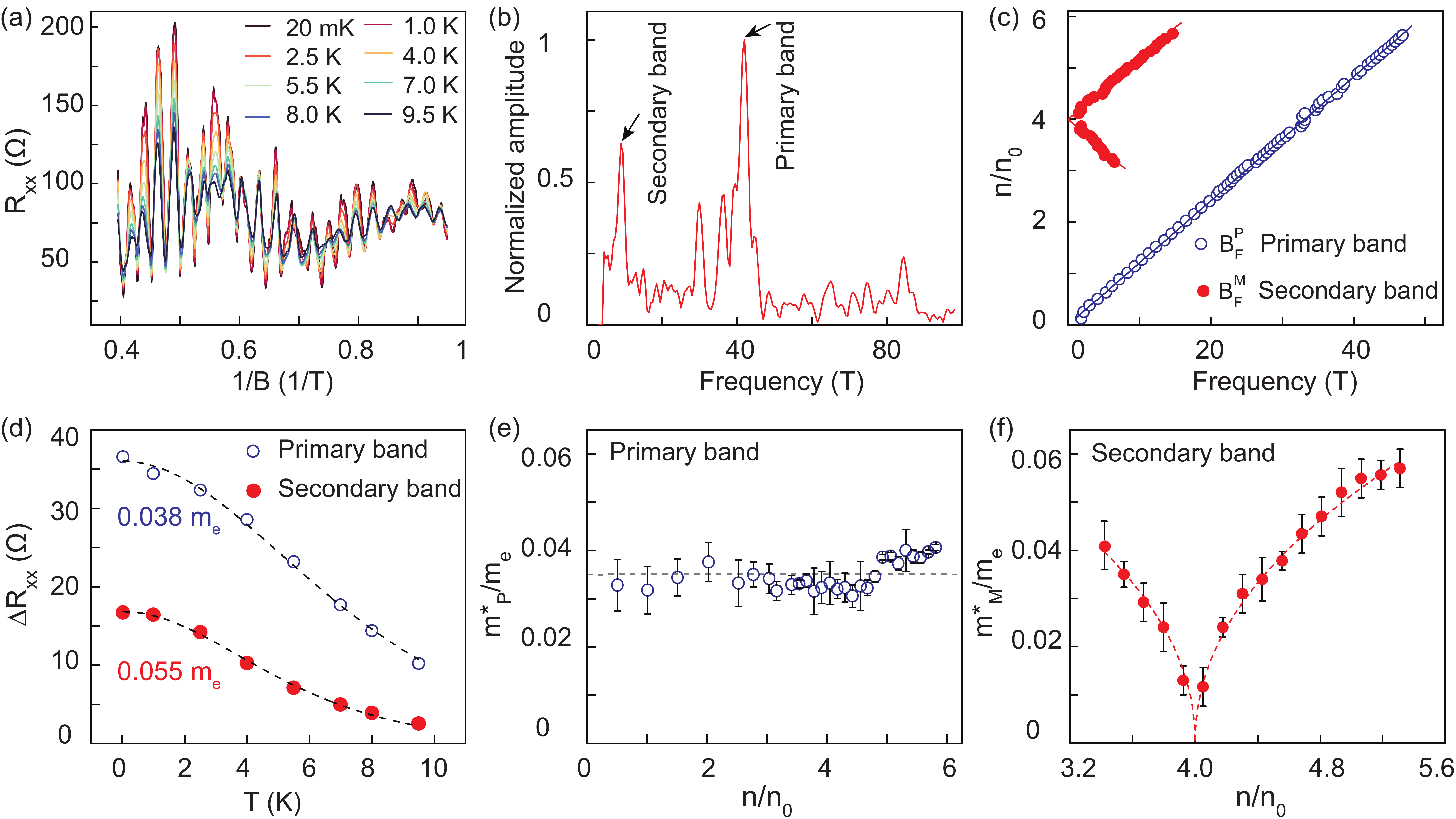}
	\caption{\textbf{Effective mass extracted from Shubnikov de Haas oscillations.}
		(a) Temperature-dependent SdH oscillations in $R_{xx}$ plotted versus inverse magnetic field $(1/B)$ at
		$n = 4.05\times10^{16}~\mathrm{m^{-2}}$ ($n/n_0 = 5.06$).
		(b) FFT of the oscillations at $T = 20~\mathrm{mK}$ showing two distinct frequency components corresponding to the primary and secondary bands.
		(c) Extracted oscillation frequencies $B_F^{P}$ (blue) and $B_F^{M}$ (red) as a function of normalized carrier density $n/n_0$; solid lines are linear fits using $B_F^{P} = |n|h/(g_s g_v e)$ and $B_F^{M} = |n - 4n_0|h/(g_s g_v e)$ with $g_s g_v = 4$.
		(d) Temperature dependence of SdH amplitude for the primary (open blue) and secondary (filled red) bands; dashed lines show Lifshitz–Kosevich fits.
		(e) Normalized effective mass of the primary band $m^*_P/m_e$ as a function of $n/n_0$, with an average value $m^*_P/m_e = 0.035$.
		(f) Normalized effective mass of the secondary band $m^*_M/m_e$ versus $n/n_0$, fitted with $m^*_M/m_e = A_M |(n - 4n_0)/n_0|^{\alpha_M}$ (red dashed line).
		Error bars in (e,f) represent standard deviations from LK fits over multiple field windows.}
	\label{fig:fig3}
\end{figure}

\clearpage

	%%% Figure 4
\begin{figure}[t]
	\includegraphics[width=0.6\columnwidth]{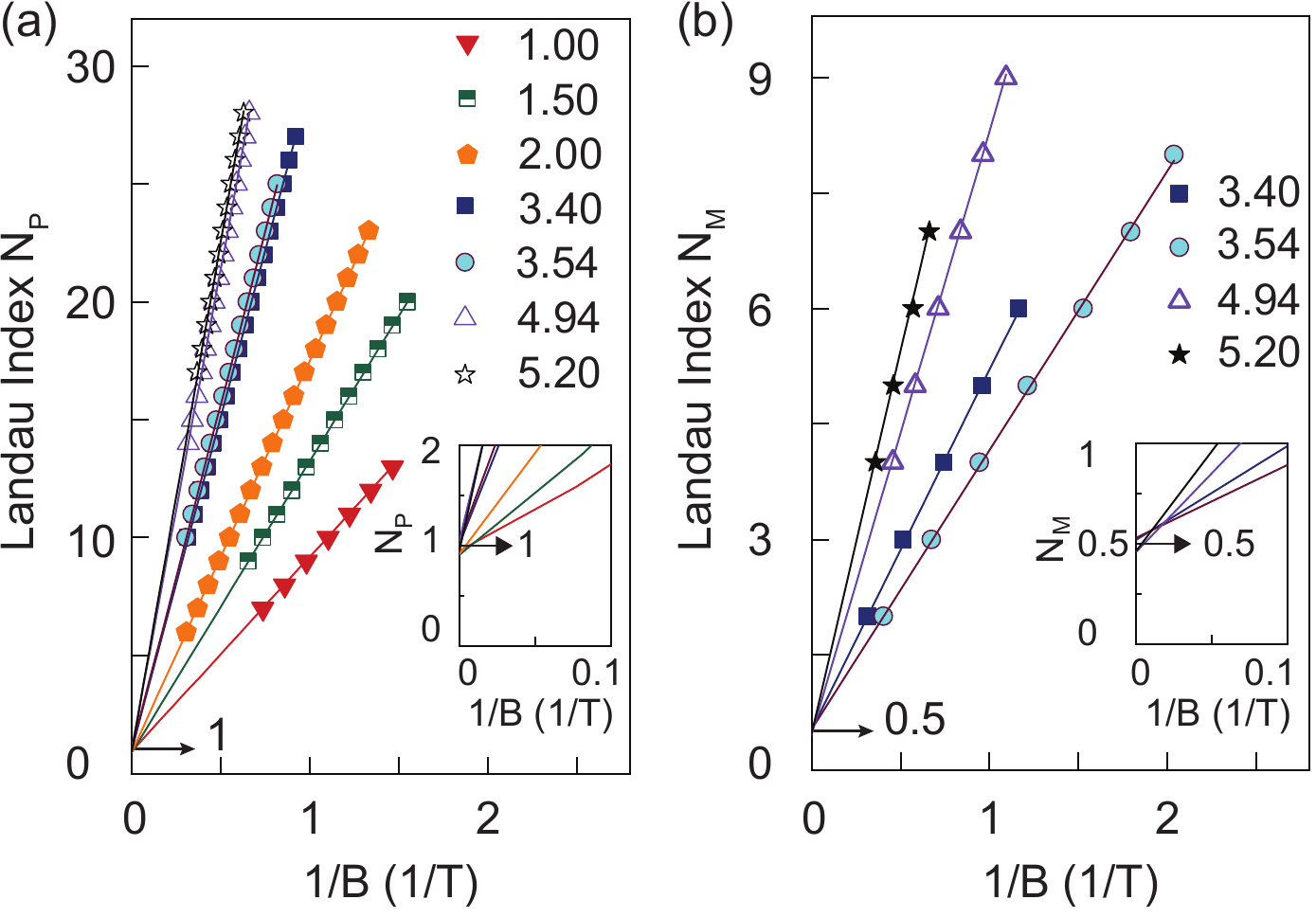}
	\caption{\textbf{Berry-phase analysis of primary and moiré-induced bands.}
		(a) Landau index $N_P$ versus inverse magnetic field $(1/B)$ for the primary BLG band for different values of $n/n_0$, showing an intercept $\beta = 1$ corresponding to a Berry phase $\Phi_B = 2\pi$.
		Inset: highlighting the linear extrapolation used to extract $\beta = 1$.
		(b) Landau index $N_M$ versus $(1/B)$ for the moiré-induced secondary band, yielding $\beta = 0.5$ ($\Phi_B = \pi$), characteristic of massless Dirac fermions. Inset: emphasizing the $\beta = 0.5$ intercept.
		The contrasting intercepts between the two bands directly confirm a transition from massive to massless quasiparticles induced by the moiré superlattice.}
	\label{fig:fig4}
\end{figure}

\clearpage

\section*{Supplementary Materials}

\renewcommand{\theequation}{S\arabic{equation}}
\renewcommand{\thesection}{S\arabic{section}}
\renewcommand{\thefigure}{S\arabic{figure}}
\renewcommand{\thetable}{S\arabic{table}}
\setcounter{table}{0}
\setcounter{figure}{0}
\setcounter{equation}{0}
\setcounter{section}{0}

	\section{\textbf{Device fabrication}}
\begin{figure*}[t] \includegraphics[width=0.55\columnwidth]{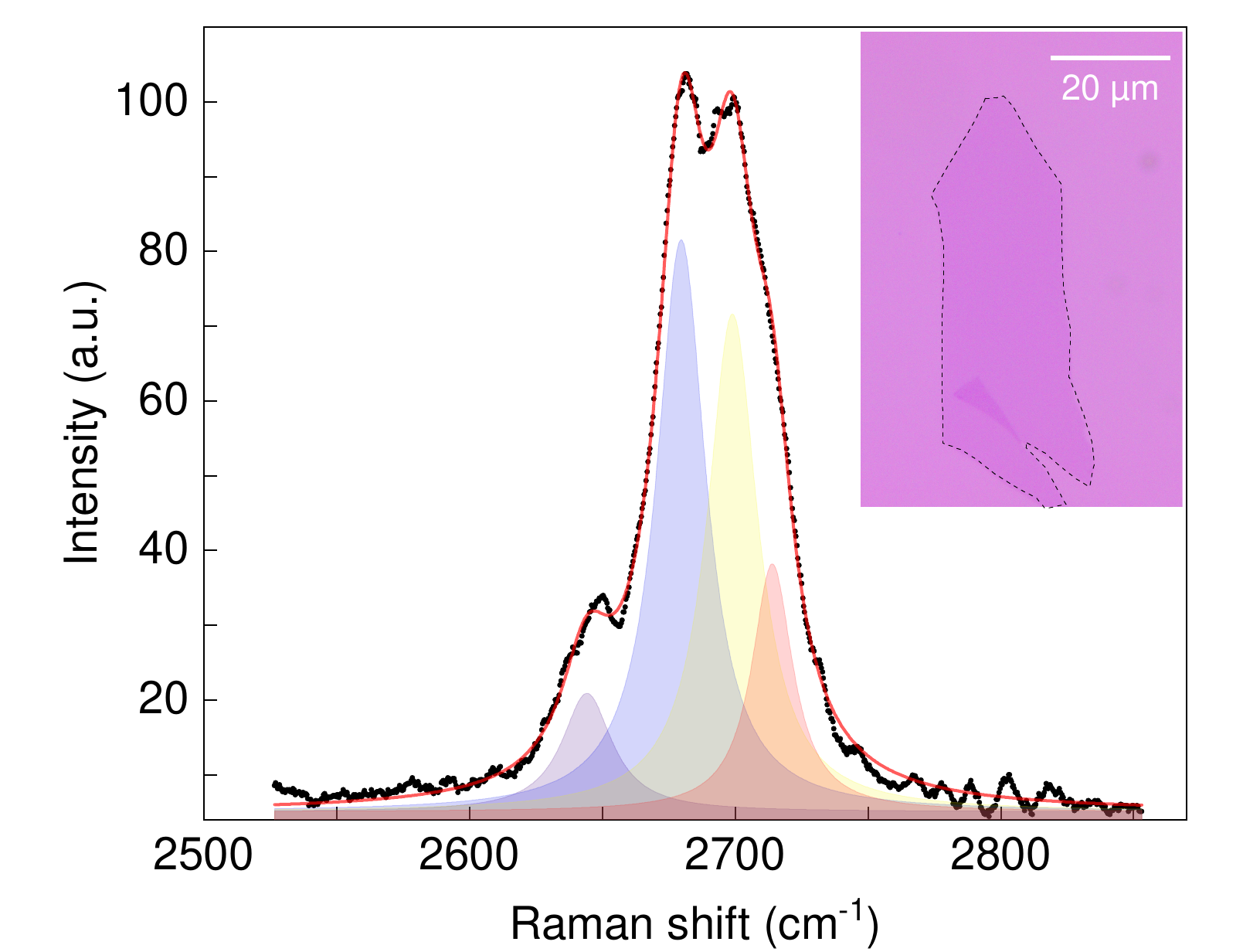} \small{\caption{ \textbf{Raman spectra of BLG flakes}. Plots of the $\mathrm{2D}$ Raman peak of the bilayer graphene used to fabricate the device. The black-filled circles are the experimentally measured Raman spectra. The red solid line is the cumulative of the four Lorentzians fitted to it; the four Lorentzians are also individually shown. Inset: Optical images of the BLG flake -- the area marked by the black line marks the portion used for device fabrication. Scale bar: 20~$\mu$m.}
		\label{fig:figS1}}
\end{figure*}

We fabricated dual-graphite-gated, hexagonal boron nitride (hBN)-encapsulated bilayer graphene devices (\textbf{D1, D2}) using a dry transfer technique~\cite{Pizzocchero2016,wang2013one, Jat2024}. Thin flakes of hBN, bilayer graphene, and graphite were mechanically exfoliated onto Si/SiO\textsubscript{2} substrates using the standard scotch tape method. Flakes were initially selected based on their optical contrast. Surface cleanliness, absence of cracks, and layer non-uniformity were evaluated under dark-field optical microscopy. The Bernal bilayer nature of the graphene was confirmed through Raman spectroscopy, as shown in Fig.\ref{fig:figS1}. The thickness and uniformity of the hBN layers were characterised using atomic force microscopy (AFM), and flakes with thicknesses in the range of 30--35~nm were chosen for device assembly.

First, a sharp-edged top hBN flake was picked up using a PDMS/PC film at $130^{\circ}\mathrm{C}$ while observing under an optical microscope. Next, a bilayer graphene flake with a well defined sharp-edge was picked up and precisely aligned with the edge of the top hBN, this alignment facilitates the moir\'{e} formation. Sequentially, a bottom hBN flake was picked up with its edge deliberately misaligned to prevent the formation of a moiré pattern with the bilayer graphene. Finally, a graphite flake serving as the bottom gate was picked up, and the entire stack was transferred onto a clean Si/SiO$_2$ wafer by melting PC at $180^{\circ}\mathrm{C}$.

Electrical contacts to the bilayer graphene were patterned using e-beam lithography, followed by reactive ion etching with a mixture of CHF$_3$/O$_2$ (40sccm /10sccm) and deposition of Cr/Au (3nm/45nm) to form one-dimensional edge contacts. The heterostructure was then etched into a Hall bar geometry. Top graphite and hBN layers were subsequently picked up using the same procedure described above and transferred onto the etched device. The top gate was electrically contacted with Cr/Au. In this device geometry, the one-dimensional Cr/Au contacts are encapsulated between the top and bottom graphite gates. This configuration ensures that the metal-graphene interface resides within a clean, hBN-encapsulated, and electrostatically screened environment with uniform gating. As a result, charge inhomogeneities and pn-junction formation at the contact edge are suppressed, {leading to cleaner SDH oscillations.} Additionally, dual gate architecture allows independent control over both the carrier density($n$) and perpendicular electric field($D$).

\section{\textbf{Twist angle estimation}}

\begin{figure}
	\includegraphics[width=\columnwidth]{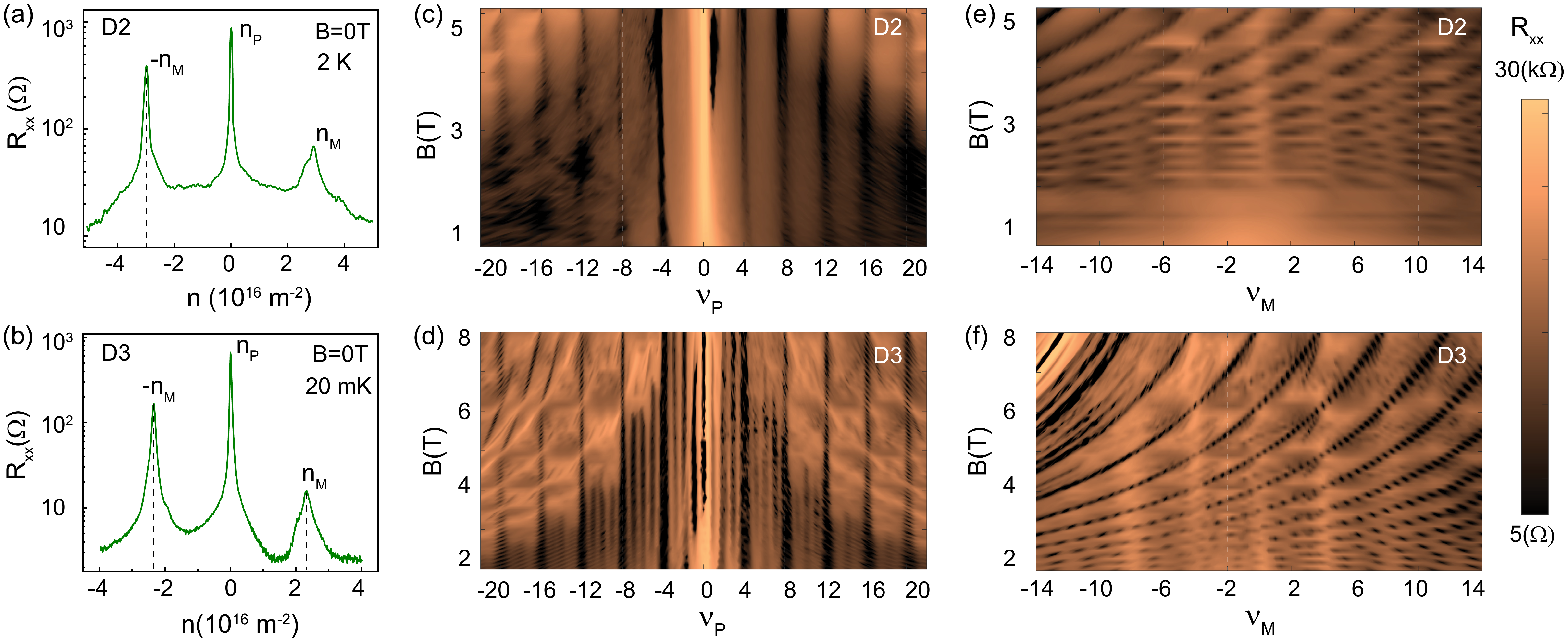}
	\caption{\small
		{\textbf{Characterization and Quantum oscillations showing Dirac band signature in Device D2 and D3}
			(a,b) Plot of ${R_{xx}}$  as a function of $n$ measured at $B = 0$~T for device \textbf{D2} ($\mathrm{T=2~K}$) and device \textbf{D3} ($\mathrm{T=20~mK}$). The primary charge neutrality peak is marked as $n_{P} = 0$. The moir\'{e} satellite peaks at $n_{M} = \pm 3\times 10^{16}\mathrm{m^{-2}}$ for \textbf{D2} and $n_{M} = \pm 2.3\times 10^{16}\mathrm{m^{-2}}$ for \textbf{D3} are marked with a dashed black line. (c,d) 2-D map of $R_{xx}$ as a function of filling fraction $\nu_\mathrm{P} = nh/eB$ and $B$ of the primary band for \textbf{D2} and \textbf{D3}, respectively. White dashed lines mark prominent minima at $\nu_\mathrm{P} = 4m$ ($m \in \mathbb{Z}$), characteristic of bilayer graphene. (e,f) 2-D map of $R_{xx}$ as a function of the effective filling fraction of secondary band $\nu_\mathrm{M} = (n - 4n_{0})h/eB$ and $B$  for \textbf{D2} and \textbf{D3}, respectively. White dashed lines highlight minima at $\nu_\mathrm{M} = 4m + 2$, indicative of Dirac-like Landau level structure. The color scale of resistance shown on the right is identical for panels (c-f).}}
	\label{fig:figS2}
\end{figure}

In the main text, Fig. 1(b) shows the plot of longitudinal resistance R$_{xx}$ versus carrier density, measured at $ T=2$ K, for the device \textbf{D1}. The resistance peak at $n=0$ originates from the charge neutrality point corresponding to the primary band. The resistance peaks at $n_{M}= \pm3.2\times 10^{16} \mathrm{m^{-2}}$ are a consequence of moir\'{e} induced band reconstruction, confirming the presence of a superlattice. Similarly, data of the device \textbf{D2} {and \textbf{D3}} is shown in Fig.\ref{fig:figS2}(a) {and (b)}  with corresponding moir\'{e} induced resistance peaks marked at $n_{M}= \pm3\times 10^{16} \mathrm{m^{-2}}$ { and $n_{M}= \pm2.3\times 10^{16} \mathrm{m^{-2}}$}, respectively.

In a periodic potential arising from a moir\'{e} superlattice, the electronic dispersion of graphene exhibits gaps at a characteristic carrier density determined by the superlattice geometry. For a superlattice unit cell of area $A$, the carrier density at which the first mini band is fully filled is given by $n_M = \frac{g_s g_v}{A}$, where $g_s = g_v = 2$ are the spin and valley degeneracy factors of graphene. In the triangular moir\'{e} pattern with wavelength $\lambda$, the area of the moir\'{e} unit cell is expressed as $A = \frac{\sqrt{3} \lambda^2}{2}$. Combining these expressions gives the relation~\cite{Jat2024,Jat202Natcomm}:
\begin{eqnarray}
	\lambda^2 = \frac{8}{\sqrt{3} n_M}  \label{Eqn:moirelength}
\end{eqnarray}
From the experimentally determined $n_M$, we extract corresponding moir\'{e} wavelength {$\lambda = 12~\mathrm{nm},12.4~\mathrm{nm}$,and $14~\mathrm{nm}$ for device \textbf{D1, D2} and \textbf{D3}}, respectively.  The relative twist angle $\theta$ between bilayer graphene (BLG) and hexagonal boron nitride (hBN) is subsequently extracted using the geometric relation~\cite{Yankowitz2012, PhysRevB.90.155406}:
\begin{eqnarray}
	\mathrm{\lambda = \frac{(1+\epsilon)a}{[\epsilon^2 +2(1+\epsilon)(1-cos(\theta))]^{1/2}}}\label{Eqn:moireangle}
\end{eqnarray}
where $a = 0.246~\mathrm{nm}$ is the graphene lattice constant, $\epsilon = 0.018$ denotes the lattice mismatch between hBN and graphene, and $\lambda$ is the moir\'{e} wavelength. Substituting {$\lambda = 12~\mathrm{nm} ~$,$~ 12.4~\mathrm{nm}$ and$~ 14~\mathrm{nm}$ into Eq.~\ref{Eqn:moireangle} for devices \textbf{D1, D2} and \textbf{D3} yields twist angle of $\theta_M = 0.60^\circ ~$,$~ 0.55^\circ$ and$~ 0^\circ$} respectively, between the BLG and hBN layer.

\section{\textbf{Relation between moir\'{e} wavelength and BZ frequency}}

Brown--Zak (BZ) oscillations are magnetotransport oscillations that arise when a rational fraction of the magnetic flux quantum $\phi_0 = h/e$ threads a superlattice unit cell. In our bilayer graphene (BLG)--hBN moir\'{e} device, $G_{xx}$ versus $1/B$ at $T = 100$~K (Fig.~1(d) of the main text) clearly shows these oscillations for device \textbf{D1}. At such high temperatures, Landau levels are thermally smeared, leaving only the BZ oscillations, whose periodicity in $1/B$ is independent of carrier density. Unlike Shubnikov--de Haas oscillations, BZ oscillations persist at elevated temperatures because they originate from recurrent Bloch states rather than quantized Landau levels.

Fast Fourier transform analysis of the oscillations gives a single frequency $f_{BZ} = 33.2$~T. This single-component spectrum rules out the presence of a supermoir\'{e} potential hBN~\cite{Jat202Natcomm,doi:10.1126/sciadv.aay8897}. The frequency $f_{BZ}$ is related to the real-space area $A$ of the moir\'{e} unit cell via $f_{BZ} = \phi_0 / A$~\cite{doi:10.1126/science.aal3357,doi:10.1073/pnas.1804572115,Jat202Natcomm,Jat2024}. Using the relation $A = \frac{\sqrt{3} \lambda^2}{2}$, we obtain $\lambda = \left( \frac{2h}{\sqrt{3} e f_{BZ}} \right)^{1/2}=12$~nm for $f_{BZ} = 33.2$~T. This value is in excellent agreement with the moir\'{e} wavelength independently determined from resistance--carrier density measurements, corresponding to a twist angle of $\theta_M = 0.60^\circ$ between BLG and hBN.

%%% Figure 2

\section{\textbf{Data from device D2 and D3}}
{Fig.~\ref{fig:figS2}(c) and (d)} shows the magnetoresistance oscillation map of the primary band as a function of the Landau index filling factor $\nu_P=nh/eB$ and magnetic field $B$ {for device \textbf{D2} and \textbf{D3}, respectively}, clearly highlighting the resistance minima position at $\nu_P=4m$ ($m\in\mathbb{Z} $), characteristic of bilayer graphene massive quasiparticles. { Fig.\ref{fig:figS2}(e) and (f)} shows the magnetoresistance oscillation of the moir\'{e} induced secondary band as a function of $\nu_M=(n-4n_0)h/eB$ and $B$ {for device \textbf{D2} and \textbf{D3}, respectively}. Resistance minima position for moir\'{e} reconstructed bands occur at $\nu_M=4(m+1/2)$ ($m\in\mathbb{Z} $), similar to the massless Dirac spectrum of single-layer graphene. These results clearly highlight that the primary band ($\nu_P=4m$) and the moir\'{e}-induced secondary band ($\nu_M=4m+2$ ) have different topologies.

\section{\textbf{Deriving electronic dispersion from the empirical effective-mass model}}

The effective mass $m^*$ is defined through the curvature of the energy band as
\begin{equation}
	\frac{1}{m^*} = \frac{1}{\hbar^2} \frac{\partial^2 E}{\partial k^2},
	\label{Eqn:effectivemass}
\end{equation}
where $E$ is the energy and $k$ is the wave vector.
For an isotropic dispersion, the $k$-space area is given by $S = \pi k^2$.
Expressing Eq.~\ref{Eqn:effectivemass} in terms of $S$ gives
\begin{equation}
	m^* = \frac{\hbar^2}{2\pi} \frac{\mathrm{d}S}{\mathrm{d}E}.
	\label{Eqn:effectivemass2}
\end{equation}

In many 2D materials, the effective mass can be empirically expressed as a function of the carrier density $n$~\cite{Novoselov2005, Tiwari2022}:
\begin{equation}
	\frac{m^*}{m_e} = A\left(\frac{n}{n_0}\right)^\alpha,
	\label{Eqn:empiricalrelation}
\end{equation}
where $m_e$ is the free-electron mass, $A$ and $\alpha$ are fitting parameters, and $n_0$ is the reference carrier density, defined as the number of carriers required to occupy one state per spin and valley in a moir\'{e} unit cell. For a fixed moir\'{e} wavelength, $n_0$ is constant; in our sample, $n_0 = 8 \times 10^{15}~\mathrm{m}^{-2}$.

Comparing Eqns.~\ref{Eqn:effectivemass2} and \ref{Eqn:empiricalrelation} yields
\begin{equation}
	\frac{\hbar^2}{2\pi} \frac{\mathrm{d}S}{\mathrm{d}E} = Am_e\left(\frac{n}{n_0}\right)^\alpha.
	\label{Eqn:empiricalrelation1}
\end{equation}

For a 2D system, the carrier density is related to the Fermi surface area by
\begin{equation}
	n = \frac{g_sg_v\,S(k)}{4\pi^2},
	\label{Eqn:ndensity}
\end{equation}
where $g_sg_v$ is the degeneracy factor. Substituting Eq.~\ref{Eqn:ndensity} into Eq.~\ref{Eqn:empiricalrelation1} gives
\begin{equation}
	\frac{\mathrm{d}E}{\mathrm{d}S} = \frac{\hbar^2}{2\pi Am_e}
	\left(\frac{4\pi^2n_0}{g_sg_v}\right)^\alpha S^{-\alpha}.
\end{equation}

Integrating with respect to $S$, we obtain
\begin{equation}
	E(S) = \frac{\hbar^2}{2\pi Am_e}
	\left(\frac{4\pi^2 n_0}{g_sg_v}\right)^\alpha
	\frac{S^{1-\alpha}}{1-\alpha} + E_0,
\end{equation}
where $E_0$ is the integration constant.
Using $S = \pi k^2$, the dispersion relation becomes
\begin{equation}
	E(k) = \frac{\hbar^2}{2Am_e(1-\alpha)}
	\left(\frac{4\pi n_0}{g_sg_v}\right)^\alpha
	k^{2(1-\alpha)} + E_0.
	\label{Eqn:dispersion}
\end{equation}
for a spin and valley degenerate bilayer graphene $g_s=g_v=2$
\begin{equation}
	E(k) = \frac{\hbar^2(\pi n_0)^\alpha}{2Am_e(1-\alpha)}
	k^{2(1-\alpha)} + E_0.
	\label{Eqn:dispersion_final}
\end{equation}

Equation~\ref{Eqn:dispersion_final} represents the energy--momentum relation.
The parameter $\alpha$ determines the band curvature:  for $\alpha = 0$, the dispersion is parabolic, whereas $\alpha = 0.5$ corresponds to a linear (Dirac-like) dispersion.

\section{\textbf{Decoupling primary and secondary bands contribution using FFT band pass filter}}

% figure 3

\begin{figure}[t]
	\includegraphics[width=\columnwidth]{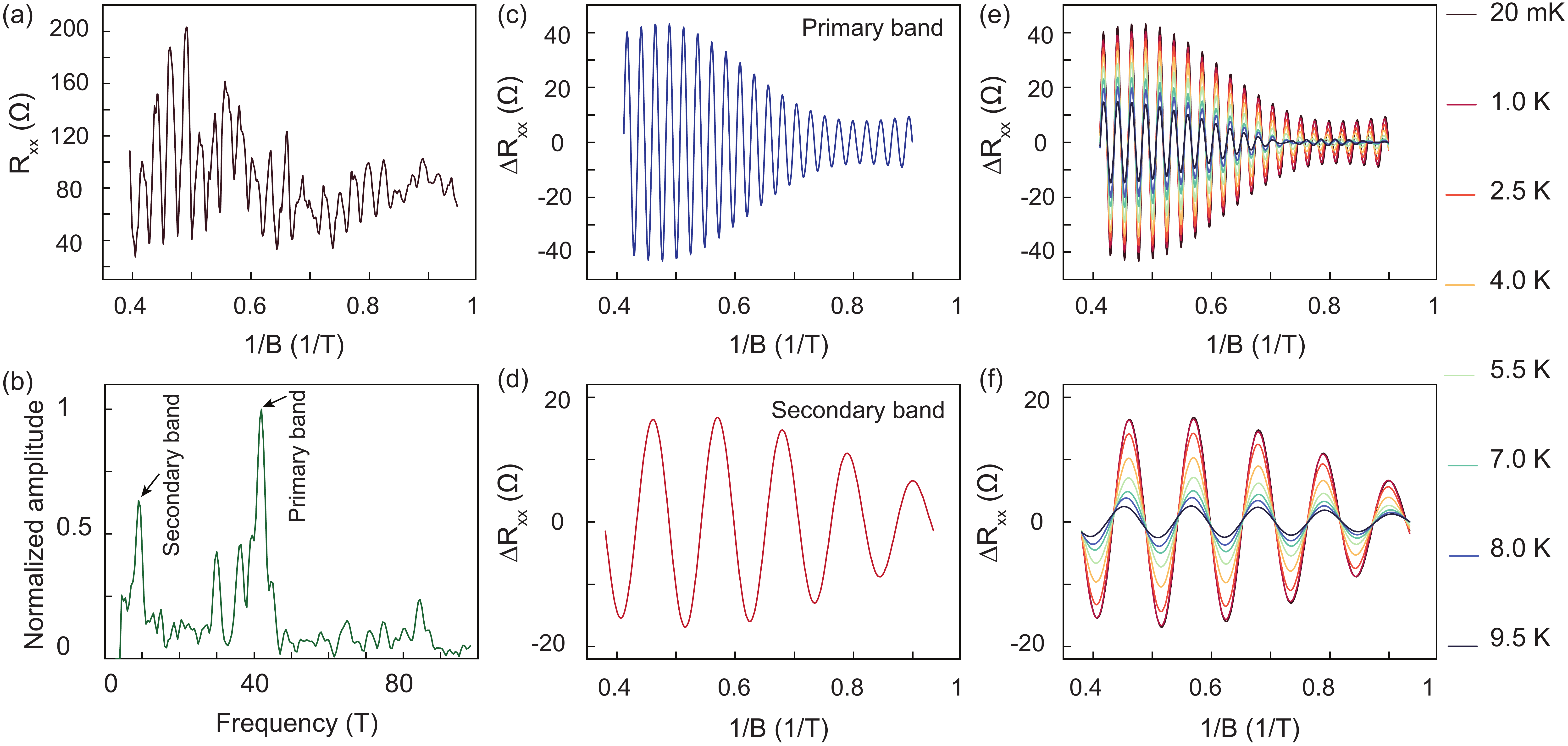}
	\caption{\small\textbf{{Decoupling the primary and secondary band contributions.}}
		(a) SdH oscillations as a function of inverse magnetic field ($1/B$) at $n = 4.05 \times 10^{16}~\mathrm{m}^{-2}$ ($n/n_0 = 5.06$) and $T = 20~\mathrm{mK}$.
		(b) FFT spectrum of the oscillations, showing two dominant frequencies corresponding to the primary ($42~\mathrm{T}$) and secondary ($9~\mathrm{T}$) bands, marked by arrows.
		(c) Filtered oscillations of the primary band using a band-pass window between $40$ and $44~\mathrm{T}$.
		(d) Filtered oscillations of the moir\'{e}-induced secondary band using a band-pass window between $7$ and $11~\mathrm{T}$.
		(e) and (f) Temperature dependence of the filtered SdH oscillations for the primary and secondary bands, respectively. The identical temperature colour scale of (e) and (f) is shown on the right.}
	\label{fig:figs3}
\end{figure}

Fig.~\ref{fig:figs3}(a) shows the measured Shubnikov–de Haas (SdH) oscillations of the longitudinal resistance $R_{xx}$ as a function of magnetic field ($1/B$)  at carrier density $n = 4.05 \times 10^{16}~\mathrm{m}^{-2}$ ($n/n_0 = 5.06$) and temperature $T = 20~\mathrm{mK}$ in device \textbf{D1}, clearly indicating presence of multiple frequencies.

To identify the underlying frequencies of these oscillations, we performed a fast Fourier transform (FFT), as shown in Fig.~\ref{fig:figs3}(b). The FFT spectrum reveals two dominant peaks, corresponding to frequencies of $42~\mathrm{T}$ and $9~\mathrm{T}$. These peaks originate from two different sets of Fermi surface pockets: the primary band at $42~\mathrm{T}$, and moir\'{e} induced secondary band at $9~\mathrm{T}$. To decouple the contribution of each band, we applied band-pass filters to isolate their respective oscillations. For the primary band, we chose a window between $40$ and $44~\mathrm{T}$, which effectively captures the $42~\mathrm{T}$ component while eliminating all other contributions. The resulting filtered oscillations are shown in Fig.~\ref{fig:figs3}(c). Similarly, for the secondary band, we applied a band-pass filter between $7$ and $11~\mathrm{T}$. This isolates the $9~\mathrm{T}$ low-frequency oscillations associated with the secondary band, as shown in Fig.~\ref{fig:figs3}(d).

Finally, we studied the temperature dependence of these filtered oscillations. Fig.~\ref{fig:figs3}(e) and Fig.~\ref{fig:figs3}(f) show the data for the primary and secondary bands, respectively. Both sets of oscillations show a decay in amplitude as the temperature increases, consistent with the thermal smearing of Landau levels and following Lifshitz–Kosevich damping. This analysis allows us to extract effective masses for each band, which are discussed in the main manuscript.

\section{\textbf{Quantum Oscillation around $n/n_0 = -4$}}

\begin{figure}[t]
	\includegraphics[width=\columnwidth]{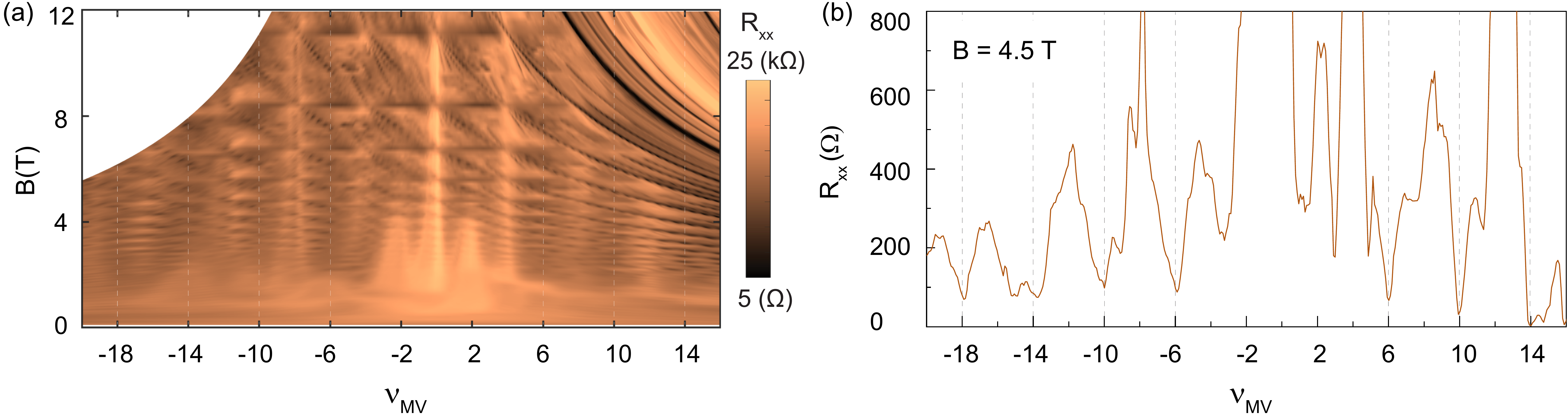}
	\caption{\small
		\textbf{Quantum oscillations revealing moir\'{e}-induced Dirac fermions on valence side secondary miniband.}
		{(a) 2-D map of $R_{xx}$ as a function of $B$ and the effective filling fraction of secondary mini band on valence side $\nu_\mathrm{MV} = (n + 4n_{0})h/eB$. White dashed lines highlight minima at $\nu_\mathrm{MV} = 4m + 2$, indicative of Dirac-like Landau level structure. (b) Line cut of panel (a) at $B = 4.5$~T, showing clear $R_{xx}$ minima at $\nu_\mathrm{MV} = 4m + 2$ for $|m| >0$.}}
	\label{fig:figS4}
\end{figure}

{The Landau fans emanating from the valence side ($n< 0$) secondary miniband corresponds to $s=-4$ in the Diophantine framework, with $t=\nu_{MV} = {(n+4n_0)h}/{eB}$. The resistance minima now occur at $\nu_{MV} = 4(m+1/2)$ (Fig.~\ref{fig:figS4}(a)).  (Fig.~\ref{fig:figS4}(b)) shows line cut of this 2D plot at $B = 4.5$~T, highligting clear $R_{xx}$ minima at $\nu_\mathrm{MV} = 4m + 2$ for $|m| >0$. A half-integer shift in the Landau-level index sequence at $s =-4$ shows the emergence of massless Dirac quasiparticles from the valence side of the secondary miniband.~\cite{Novoselov2005, Zhang2005}.}

%%% Figure 4

\section{\textbf{Comparing conduction ($n>0)$ and valence band ($n<0$) SDH oscillation}}

\begin{figure}[t]
	\includegraphics[width=0.95\columnwidth]{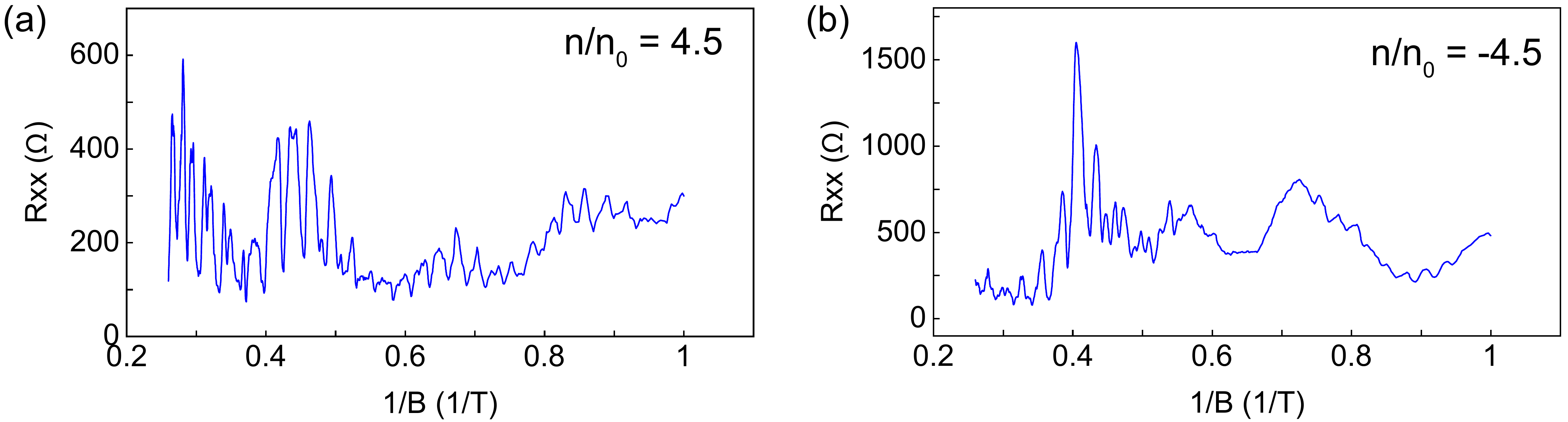}
	\caption{\small
		\textbf{Quantum oscillations in the conduction and valence bands.}
		{(a) Shubnikov-de Haas (SdH) oscillations in the longitudinal resistance $R_{xx}$ plotted as a function of inverse magnetic field $(1/B)$ for the conduction band at carrier density $n = 3.6\times10^{16}~\mathrm{m^{-2}}$ ($n/n_0 = 4.5$).
			(b) SdH oscillations in $R_{xx}$ plotted versus $(1/B)$ for the valence band at $n = -3.6\times10^{16}~\mathrm{m^{-2}}$ ($n/n_0 = -4.5$). Exhibiting clear oscillations for the conduction band and considerably weaker oscillations for the valence band.
	}}
	\label{fig:figS5}
\end{figure}

{Fig.~\ref{fig:figS5}(a) shows the clean Shubnikov-de Hass oscillations of $R_{xx}$ versus $1/B$ in the conduction band at
	$n = 3.6 \times 10^{16}\mathrm{m^{-2}}$ ($n/n_0 = 4.5$), exhibiting cleaner SdH oscillations. In contrast, the SdH oscillations measured in the valence band at $n = -3.6 \times 10^{16} \mathrm{m^{-2}}$ ($n/n_0 = -4.5$) are significantly weaker, as shown in Fig.~\ref{fig:figS5}(b). Therefore, reliable quantitative analysis is possible for the conduction band oscillations ($n> 0$), whereas the valence band oscillations ($n< 0$) are less suitable for detailed analysis.}

%%% Figure 5

\section{\textbf{Band gap estimation of the secondary minibands}}

%%% Figure S6
\begin{figure}[t]
	\includegraphics[width=0.9\columnwidth]{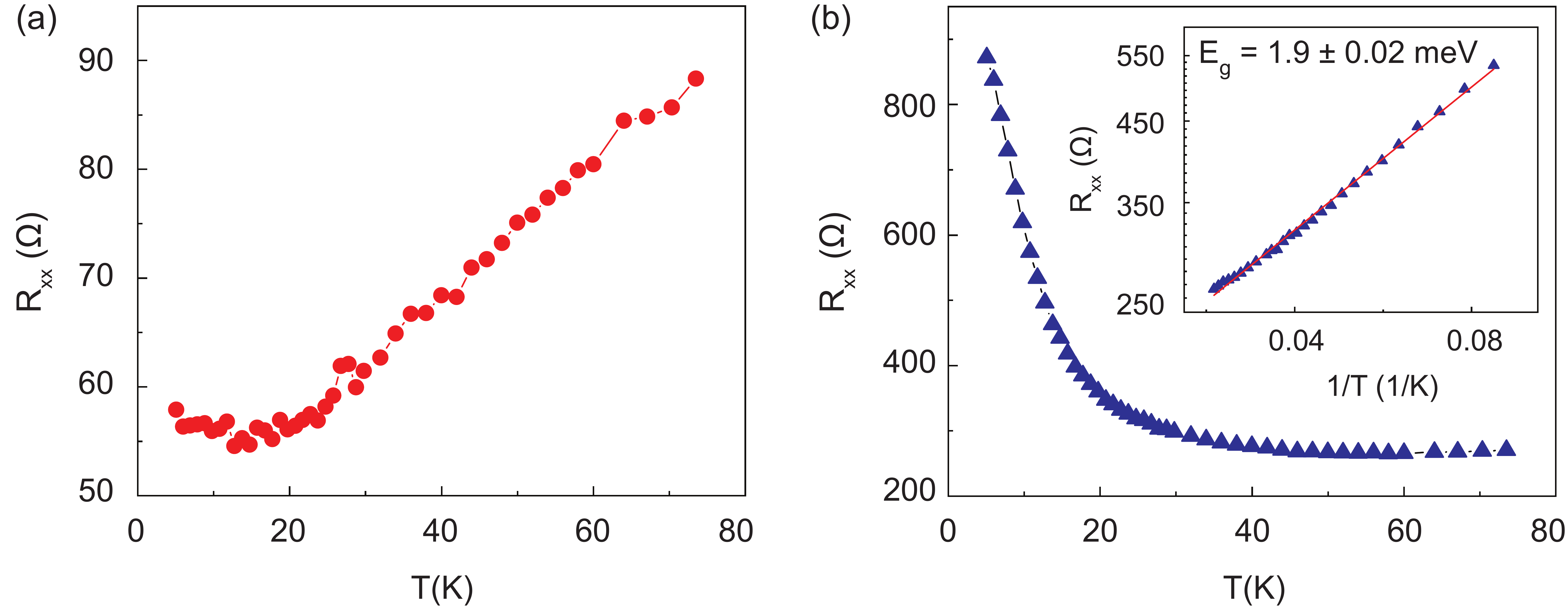}
	\caption{\small
		\textbf{Temperature dependence and band gap estimation of secondary minibands.}
		{(a) Longitudinal resistance $R_{xx}$ as a function of temperature $T$ for the secondary miniband at filling $n/n_0 = 4$, showing metallic behaviour, and no band gap.
			(b) Longitudinal resistance $R_{xx}$ as a function of temperature $T$ for the secondary miniband at $n/n_0 = -4$, exhibiting insulating behaviour.
			Inset: Arrhenius plot of $R_{xx}$ on a logarithmic scale as a function of $1/T$. The red solid line is a fit to the Arrhenius relation $R_{xx} = R_0 \exp(E_g/2k_B T)$, yielding an energy gap $E_g = 1.9\pm 0.02$~meV.}
	}
	\label{fig:figS6}
\end{figure}

{Fig.~\ref{fig:figS6} shows the temperature dependence of the longitudinal resistance $R_{xx}$ measured at fixed carrier densities corresponding to the secondary miniband fillings $n/n_0 = \pm 4$. At $n/n_0 = 4$, $R_{xx}$ increases monotonically with temperature [Fig.~\ref{fig:figS6}(a)], exhibiting metallic behaviour and the absence of an energy gap. In contrast, at $n/n_0 = -4$, $R_{xx}$ decreases monotonically with temperature [Fig.~\ref{fig:figS6}(b)], characteristic of insulating behaviour, suggesting the opening of a gap in the secondary miniband.}

{The energy gap is estimated using an Arrhenius relation,
	\begin{equation}
		R_{xx} = R_0 \exp(\frac{E_g}{2k_B T}).
	\end{equation}
	The inset of Fig.~\ref{fig:figS6}(b) shows a linear dependence of $R_{xx}$ on a logarithmic scale as a function of $1/T$, from which an energy gap of $E_g = 1.9\pm 0.02$~meV is extracted. These results show a small band gap opening only for negative filling ($E_g \approx 1.9$~meV), while the corresponding positive filling remains gapless, in agreement with the previous measurements ~\cite{Kim2018}}.

\section {\textbf{Berry phase extraction from Shubnikov-de Haas oscillations}}

{The Berry phase is extracted from Shubnikov-de Haas (SdH) oscillations of the longitudinal resistance $R_{xx}$, the oscillatory component of these oscillations follows~\cite {Novoselov2005, Zhang2005}:}
\begin{equation}
	\Delta R_{xx} \propto \cos[2\pi(\frac{B_F}{B} + \frac{1}{2} + \beta)]
\end{equation}

{where $B_F$ is the oscillation frequency and $2\pi\beta$ is the corresponding Berry phase. Landau level (LL) indices $N$ are assigned to the minima of $R_{xx}$, giving $\mathrm{N} = \frac{B_F}{B} + \beta$.
	A linear fit of $\mathrm{N}$ versus $1/B$ gives the intercept $\beta$.}

\begin{figure}[t]
	\includegraphics[width=\columnwidth]{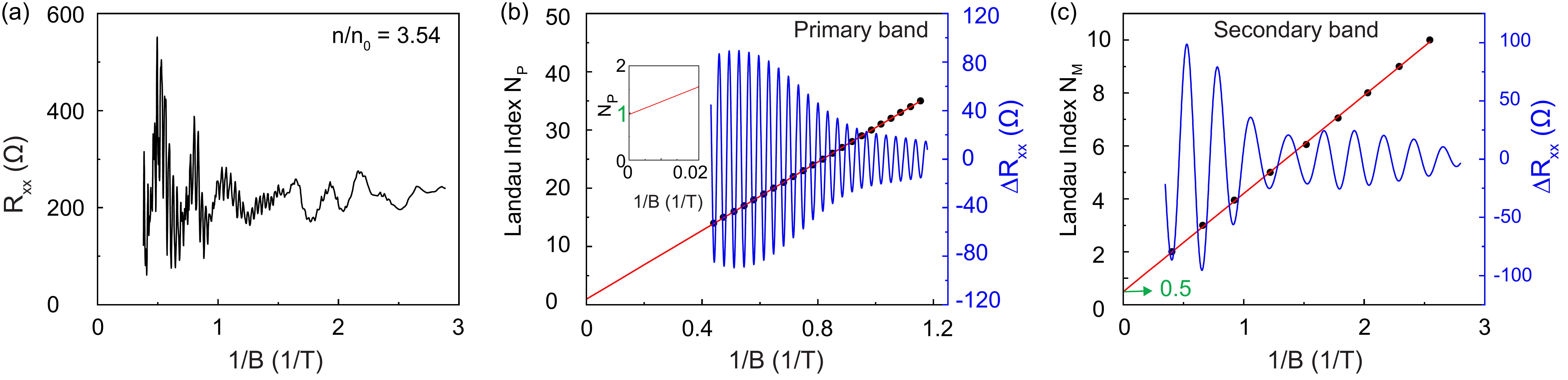}
	\caption{\small
		\textbf{ Berry-phase extraction from SdH oscillations.}
		{(a) Longitudinal resistance ($R_{xx}$) as a function of inverse magnetic field ($1/B$) at $n/n_0 = 3.54$.
			(b) Frequency-decoupled Shubnikov-de Haas oscillations of the primary band with the corresponding Landau-level (LL) index $\mathrm{N_P}$ assigned to the minima of $R_{xx}$, yielding an intercept $\beta = 0.97 \pm 0.04$ ($\Phi_B = 2\pi$); the inset shows the extrapolation of the Landau index to $1/B = 0$. (c) Frequency-decoupled oscillations and Landau index $\mathrm{N_M}$ versus $1/B$ for the secondary band, giving intercept $\beta= 0.49 \pm 0.02$ ($\Phi_B = \pi$).}
	}
	\label{fig:figS7}
\end{figure}

{Fig.~\ref{fig:figS7}(a) shows the SdH oscillations at $n/n_0 = 3.54$, containing contributions from both the primary band and a moir\'{e}-induced secondary band. Using the FFT decoupling procedure described in Section~S6, SdH components are extracted independently for each band. Fig.~\ref{fig:figS7}(b), shows the primary band SdH component and extracted intercept of $\beta = 0.97 \pm 0.04$, corresponding to a Berry phase of $2\pi$, while the secondary band component shown in Fig.~\ref{fig:figS7}(c), gives $\beta = 0.49 \pm 0.02$, resulting in a Berry phase of $\pi$. The extracted intercepts remain stable within $\pm 0.05$ upon varying the fitting window and background subtraction.}

%%% Figure S7
 \clearpage

\textbf{References}

	\clearpage

\end{document}